\documentclass{llncs}

\usepackage{amsmath,cite,graphicx,pslatex}
\usepackage[loose]{subfigure}

\newlength{\gnuplotFigWidth}
\begin{document}

\newcommand{\abb}[1] {$\mbox{#1}$}

\newcommand{\ack}     {\abb{ACK}}
\newcommand{\cds}     {\abb{CDS}}
\newcommand{\ch}      {\abb{CH}}
\newcommand{\dsr}     {\abb{DSR}}
\newcommand{\ed}      {\abb{ED}}
\newcommand{\edpos}   {\abb{EDPOS}}
\newcommand{\egide}   {\abb{EGIDE}}
\newcommand{\gps}     {\abb{GPS}}
\newcommand{\gspr}    {\abb{GSPR}}
\newcommand{\hello}   {\abb{HELLO}}
\newcommand{\id}      {\abb{id}}
\newcommand{\inria}   {\abb{INRIA}}
\newcommand{\ip}      {\abb{IP}}
\newcommand{\ircica}  {\abb{IRCICA}}
\newcommand{\lifl}    {\abb{LIFL}}
\newcommand{\mac}     {\abb{MAC}}
\newcommand{\mpr}     {\abb{MPR}}
\newcommand{\np}      {\abb{NP}}
\newcommand{\ns}      {\abb{NS}$2$}
\newcommand{\nserc}   {\abb{NSERC}}
\newcommand{\olsr}    {\abb{OLSR}}
\newcommand{\pda}     {\abb{PDA}}
\newcommand{\pops}    {\abb{POPS}}
\newcommand{\site}    {\abb{SITE}}
\newcommand{\wifi}    {\abb{WiFi}}

\title{Maximizing the Probability of Delivery of Multipoint Relay Broadcast Protocol in Wireless\\Ad Hoc Networks with a Realistic Physical Layer}

\author{Fran\c{c}ois Ingelrest \and David Simplot-Ryl}
\institute{IRCICA/LIFL, University of Lille 1.\\CNRS UMR 8022, INRIA Futurs, France.\thanks{This work was partially supported by a grant from CPER Nord-Pas-de-Calais/FEDER TAC COM'DOM, INRIA research action IRAMUS and CNRS national platform RECAP.}\\\email{\{Francois.Ingelrest, David.Simplot\}@lifl.fr}}
\maketitle

\begin{abstract}
It is now commonly accepted that the unit disk graph used to model the physical layer in wireless networks does not reflect real radio transmissions, and that the lognormal shadowing model better suits to experimental simulations. Previous work on realistic scenarios focused on unicast, while broadcast requirements are fundamentally different and cannot be derived from unicast case. Therefore, broadcast protocols must be adapted in order to still be efficient under realistic assumptions. In this paper, we study the well-known multipoint relay protocol (\mpr{}). In the latter, each node has to choose a set of neighbors to act as relays in order to cover the whole $2$-hop neighborhood. We give experimental results showing that the original method provided to select the set of relays does not give good results with the realistic model. We also provide three new heuristics in replacement and their performances which demonstrate that they better suit to the considered model. The first one maximizes the probability of correct reception between the node and the considered relays multiplied by their coverage in the $2$-hop neighborhood. The second one replaces the coverage by the average of the probabilities of correct reception between the considered neighbor and the $2$-hop neighbors it covers. Finally, the third heuristic keeps the same concept as the second one, but tries to maximize the coverage level of the $2$-hop neighborhood: $2$-hop neighbors are still being considered as uncovered while their coverage level is not higher than a given coverage threshold, many neighbors may thus be selected to cover the same $2$-hop neighbors.
\end{abstract}

 %%%%%%%%%%%%%%%%%%%%%%%%%%%%%%%%%%%%%%%%%%%%%%%%%%%%%%%
%    ____     __              __         __  _          %
%   /  _/__  / /________  ___/ /_ ______/ /_(_)__  ___  %
%  _/ // _ \/ __/ __/ _ \/ _  / // / __/ __/ / _ \/ _ \ %
% /___/_//_/\__/_/  \___/\_,_/\_,_/\__/\__/_/\___/_//_/ %
%                                                       %
 %%%%%%%%%%%%%%%%%%%%%%%%%%%%%%%%%%%%%%%%%%%%%%%%%%%%%%%
\section{Introduction}
\label{sec:introduction}

Nowadays, wireless networking has become an indispensable technology. However, the most deployed technology, known as \wifi{}, is too restrictive, as users must stay near to fixed access points. Therefore, the latter must be sufficiently deployed and correctly configured to offer a good quality of service. Moreover, there exists some more unusual situations where an infrastructure may be unavailable (\emph{e.g.}, rescue areas). The future of this technology probably lies in wireless ad hoc networks, which are designed to be functional without any infrastructure. They are defined to be composed of a set of mobile or static hosts operating in a self-organized and decentralized manner, which communicate together thanks to radio interfaces. Hosts may be either terminals or routers, depending on the needs of the system, leading to a cooperative multi-hop routing.

Broadcasting is one of the most important communication task in those networks, as it is used for many purposes such as route discovery (\emph{e.g.}, \olsr{} \cite{JMCLQV-OPTIMIZED-OLSRPAHN}) or synchronization. In a straightforward solution to broadcasting, hosts blindly relay packets upon first reception to their neighborhood in order to fully cover the network. However, due to known physical phenomena, this leads to the broadcast storm problem \cite{NTCS-BROADCAST-TBSPIAMAHN}. Moreover, this is a totally inefficient algorithm, because most of the retransmissions are not needed to ensure the global delivery of the packet, and a huge amount of energy is thus unnecessarily wasted. Many other algorithms have been proposed in replacement. Some of them are centralized (a global knowledge of the network is needed), while the others are localized (hosts only need to know their local neighborhood to take decisions). Obviously, the latter better fit to ad hoc networks and their decentralized architecture.

All the proposed broadcast schemes have always been studied under ideal scenario, where the unit disk graph is used to model communications between hosts. In this model, two hosts can communicate together if the distance between them is no more than a given communication radius, and packets are always received without any error. Recently, this model has been highly criticized as it does not correctly reflect the behavior of transmissions in a real environment \cite{SNK-DESIGN-DGRPAHSNRPL}. Indeed, signal strength fluctuations have a significant impact on performance, and thus cannot be ignored when designing communication protocols for ad hoc and sensor networks. Unfortunately, this has been the case until now for broadcast protocols.

In this paper, we consider the well-known multipoint relay protocol (\mpr{}) \cite{QVL-MULTIPOINT-MRFBMMWN}, used for broadcasting in ad hoc networks, under a more realistic scenario where the probability of correct reception of a packet smoothly decreases with the distance between the emitter and the receiver(s). We thus replace the unit disk graph model by the lognormal shadowing model \cite{QK-DEMAND-ODRIMTIOARPLM} to simulate a more realistic physical layer, and provide experimental results. As they demonstrate the need for a more suitable algorithm, we also propose several modifications to \mpr{} in order to maximize the delivery ratio of the broadcast packet, while minimizing the number of needed retransmissions. By experimentation, we show that these new versions are much more efficient than the original one under the considered realistic scenario.

\medskip

The remainder of this paper is organized as follows: we first provide the definitions needed by our models, while in Sec.~\ref{sec:relatedwork} a detailed description of \mpr{} is proposed. In Sec.~\ref{sec:originalmpr}, we provide an analysis of the behavior of the original algorithm used in \mpr{} with the realistic physical layer. We then describe in Sec.~\ref{sec:newmpr} new algorithms that better fit the latter. We finally conclude in Sec.~\ref{sec:conclusion} and give some directions for future work.

 %%%%%%%%%%%%%%%%%%%%%%%%%%%%%%%%%%%%%%%%%%%%%%%%%%%%
%    ___          ___       _               _        %
%   / _ \_______ / (_)_ _  (_)__  ___ _____(_)__ ___ %
%  / ___/ __/ -_) / /  ' \/ / _ \/ _ `/ __/ / -_|_-< %
% /_/  /_/  \__/_/_/_/_/_/_/_//_/\_,_/_/ /_/\__/___/ %
%                                                    %
 %%%%%%%%%%%%%%%%%%%%%%%%%%%%%%%%%%%%%%%%%%%%%%%%%%%%
\section{Preliminaries}
\label{sec:preliminaries}

The common representation of a wireless network is a graph $G=(V,E)$, where $V$ is the set of vertices (the hosts, or nodes) and $E \subseteq V^2$ the set of edges which represents the available communications: there exists an ordered pair $(u,v) \in E$ if the node $v$ is able to physically receive packets sent by $u$ (in a \emph{single-hop} fashion). The neighborhood set $\mbox{N}(u)$ of the node $u$ is defined as $\{v : (u,v) \in E \vee (v,u) \in E\}$. The density of the network is equal to the average number of nodes in a given communication area. Each node $u$ is assigned a unique identifier (this can simply be, for instance, an \ip{} or a \mac{} address).
 
We assume that nodes are aware of the existence of each neighboring node within a distance of $2$ hops (we call this a $2$-hop knowledge). In ad hoc networks, the neighborhood discovery is generally done thanks to small control (\hello{}) messages which are regularly sent by each host. A $2$-hop knowledge can easily be acquired thanks to two rounds of \hello{} exchanges: nodes can indeed insert the identifiers of their neighbors in their own beacon messages.

\medskip

In our mathematical model, the existence of a pair $(u,v) \in E$ is determined by the considered physical layer model and depends on several conditions, the most obvious one being the distance between $u$ and $v$. In the most commonly used model, known as the unit disk graph model, a bidirectional edge exists between two nodes if the distance between them is not greater than a given communication radius $R$ (it is assumed that all nodes have the same communication radius). In this model, the set $E$ is then simply defined by:

\begin{equation}
E = \{(u,v) \in V^2 \mid u \neq v \; \wedge \; \mbox{dist}(u,v) \leq R\},
\end{equation}

\noindent
$\mbox{dist}(u,v)$ being the Euclidean distance between nodes $u$ and $v$.

This model, while being well spread, cannot be considered as realistic. Indeed, it is assumed that packets are always received without any error, as long as the distance between the emitter and the receiver is smaller than the communication radius. This totally ignores random variations in the received signal strength, while it was demonstrated that their impact is really significant.

These fluctuations generate erroneous bits in the transmitted packets. If the error rate is sufficiently low, these bits can be repaired thanks to correction codes. However, if it is too high, then the packet must be dropped and a new emission must be done. This supposes the existence of an acknowledgement mechanism (\ack{} packets) that cannot be used in broadcasting tasks due to the really high number of emitters. Our work thus only relies upon the probability of correct reception, which is influenced by a lot of factors (\emph{e.g.}, power of emission, distance with the receiver(s), presence of obstacles). We suppose that all nodes have the same transmitting radius, so the power of emission does not have to be taken into account here.

To consider the signal fluctuations, we change $G$ into a weighted graph where each edge $(u,v) \in E$ holds the probability $\mbox{p}(u,v)$ of correct reception between the two nodes $u$ and $v$. To determine these probabilities, we chose to consider the lognormal shadowing model \cite{SNK-DESIGN-DGRPAHSNRPL} in our simulations. We used an approximated function $\mbox{P}(x)$ described in \cite{KNS-HOP-HCOPBPRAAHWNRPL}:

\begin{equation}
\mbox{P}(x)= \left\{
    \begin{array}{ll}
        1-\frac{(\frac{x}{R})^{2\alpha{}}}{2} & \mbox{if $0 < x \leq R$,}   \\
                                              &                             \\
        \frac{(\frac{2R-x}{R})^{2\alpha{}}}{2}& \mbox{if $R < x \leq 2R$,}  \\
                                              &                             \\
        0                                     & \mbox{otherwise,}           \\
    \end{array}
\right.
\end{equation}

\begin{figure}[t]
\centering
\includegraphics[angle=-90,width=\gnuplotFigWidth]{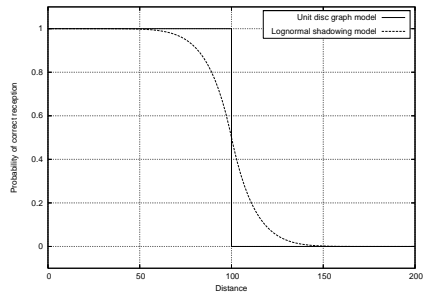}
\caption{Unit disk graph and lognormal shadowing models ($R=100$, $\alpha=4$).}
\label{fig:modelsudglns}
\end{figure}

\noindent
$\alpha{}$ being the power attenuation factor, and $x$ the considered distance. Fig.~\ref{fig:modelsudglns} illustrates this model with $R=100$ and $\alpha = 4$.

We assume that each node $u$ is able to determine the probability $\mbox{p}(u,v)$ of correct reception of a packet that would be sent to a neighbor $v$. The gain of this knowledge may be simply achieved thanks to beacon messages: based on the quantity of correctly received \hello{} packets, $v$ is able to determine an approximated value of $\mbox{p}(u,v)$. Node $v$ may then include this value in its own beacon messages.

One of the major criticisms of the unit disk graph model is that it does not model the presence of obstacles between nodes. The lognormal shadowing model neither considers them, but we argue that it is sufficient enough for simulations. The most important factor is the weighting of edges by reception probabilities, the method used to distribute the latter is not important to compare protocols in general cases. A realistic model would be mandatory to simulate existing situations and to extract exact values. But in real cases, an obstacle would decrease the probability held by the corresponding edge and would thus be detected by nodes when counting \hello{} messages (if such a method is used). This means that in those cases, the broadcast algorithm would use `real' probabilities and its behavior would be adapted to the situation.

\medskip

The two previous physical models introduce two different behaviors:

\begin{itemize}

    \item In the unit disk graph model, one has to maximize the length of each hop so that a single emission is able to reach as many mobiles as possible. The quantity of needed emitters is thus greatly reduced.

    \item In the lognormal shadowing model, maximizing the length of each hop leads to smaller probabilities of correct reception, but minimizing them leads to a lot of spent energy.

\end{itemize}

Some papers have already been published about routing in a realistic environment. Amongst them, DeCouto et al.~\cite{CABM-HIGH-AHTPMMHWR} and Draves et al.~\cite{DPZ-ROUTING-RMRMHWMN} investigate the question of routing metrics for unicast protocols in wireless networks with a realistic physical layer: the key insight in most of this work is that hop-count based shortest-path routing protocols result in transmissions over long links. While this reduces the hop-count of routes, it also decreases the received signal strength at the receiver of these links, leading to very high loss rates and low end-to-end throughput. These papers also propose other routing metrics which incorporate link-quality (\emph{e.g.}, in terms of error, congestion).

To the best of our knowledge, this paper is the first one to consider broadcasting over a realistic physical layer. Broadcast fundamentally differs from unicast, and leads to a different tradeoff between the length of each hop and the number of relays. Indeed, in a broadcast process, a node can rely on the redundancy introduced by other emitters. Further relays may thus be selected without decreasing the final delivery ratio. This is not possible in routing, as a given emitter is the only one able to transmit the packet to the next hop. The redundancy of broadcasting must be fully considered in order to improve the performance of the underlying protocol.

 %%%%%%%%%%%%%%%%%%%%%%%%%%%%%%%%%%%%%%%%%%%%%%%%%%%%%%%%%%
%    ___      __     __         __  _      __         __   %
%   / _ \___ / /__ _/ /____ ___/ / | | /| / /__  ____/ /__ %
%  / , _/ -_) / _ `/ __/ -_) _  /  | |/ |/ / _ \/ __/  '_/ %
% /_/|_|\__/_/\_,_/\__/\__/\_,_/   |__/|__/\___/_/ /_/\_\  %
%                                                          %
 %%%%%%%%%%%%%%%%%%%%%%%%%%%%%%%%%%%%%%%%%%%%%%%%%%%%%%%%%%
\section{Related Work}
\label{sec:relatedwork}

As stated in Sec.~\ref{sec:introduction}, the easiest method for broadcasting a packet is to have all nodes forward it at least once to their neighborhood: this method is known as \emph{blind flooding}. However, such a simple behavior has huge drawbacks: too many packets are lost due to collisions between neighboring nodes (this can lead to a partial coverage of the network) and far too much energy is consumed. Many other solutions have been proposed to replace it, and an extensive review of them can be found in \cite{ISS-ENERGY-EEBWMAHN}.

\medskip

Among all these solutions, we have chosen to focus on the multipoint relay protocol (\mpr{}) described in \cite{QVL-MULTIPOINT-MRFBMMWN} for several reasons:

\begin{itemize}
  \item It is efficient using the unit disk graph model.
  \item It is used in the well-known standardized routing protocol \olsr{} \cite{JMCLQV-OPTIMIZED-OLSRPAHN}.
  \item It can be used for other miscellaneous purposes (\emph{e.g.}, computing connected dominating sets \cite{AJV-COMPUTING-CCDSWMR}).
\end{itemize}

In this algorithm, it is assumed that nodes have a $2$-hop knowledge: they are aware of their neighbors ($1$-hop distance), and the neighbors of these neighbors ($2$-hop distance). Its principle is as follows. Each node $u$ that has to relay the message must first elect some of its $1$-hop neighbors to act themselves as relays in order to reach the $2$-hop neighbors of $u$. The selection is then forwarded within the packet and receivers can thus determine if they have been selected or not: each node that receives the message for the first time checks if it is designated as a relay node by the sender, and if it is the case, the message is forwarded after the selection of a new relaying set of neighbors. A variant exists where nodes proactively select their relays before having to broadcast a packet, and selection is sent within \hello{} messages.

\medskip

Obviously, the tricky part of this protocol lies in the selection of the set of relays $\mbox{MPR}(u)$ within the $1$-hop neighbors of a node $u$: the smaller this set is, the smaller the number of retransmissions is and the more efficient the broadcast is. Unfortunately, finding such a set so that it is the smallest possible one is a \np{}-complete problem, so a greedy heuristic is proposed by Qayyum et al., which can be found in \cite{L-RATIO-OROIFC}. Considering a node $u$, it can be described as follows:

\begin{enumerate}
  \item Place all $2$-hop neighbors (considering only outgoing links) in a set $\mbox{MPR}'(u)$ of uncovered $2$-hop neighbors.
  \item While there exists a $1$-hop neighbor $v$ which is the only common neighbor of $u$ and some nodes in $\mbox{MPR}'(u)$: add $v$ to $\mbox{MPR}(u)$, remove its neighbors from $\mbox{MPR}'(u)$.
  \item While the set $\mbox{MPR}'(u)$ is not empty, repeatedly choose the $1$-hop neighbor $v$ not present in $\mbox{MPR}(u)$ that covers the greatest number of nodes in $\mbox{MPR}'(u)$. Each time a new node is added to $\mbox{MPR}(u)$, remove its neighbors from $\mbox{MPR}'(u)$. In case of tie, choose the node with the highest degree.
\end{enumerate}

%There exists a variant with a fourth step, where a few `useless' relays can be removed: these are the nodes which cover the same set of $2$-hop neighbors as a set of other relays. This step increases the complexity of computation, and does not bring a really noticeable improvement. Moreover, the removal of redundant relays greatly decreases the probability of delivery using the lognormal shadowing model, this is why we chose to not consider this extra step in this paper.

\begin{figure}[t]
\centering
\includegraphics[width=100pt]{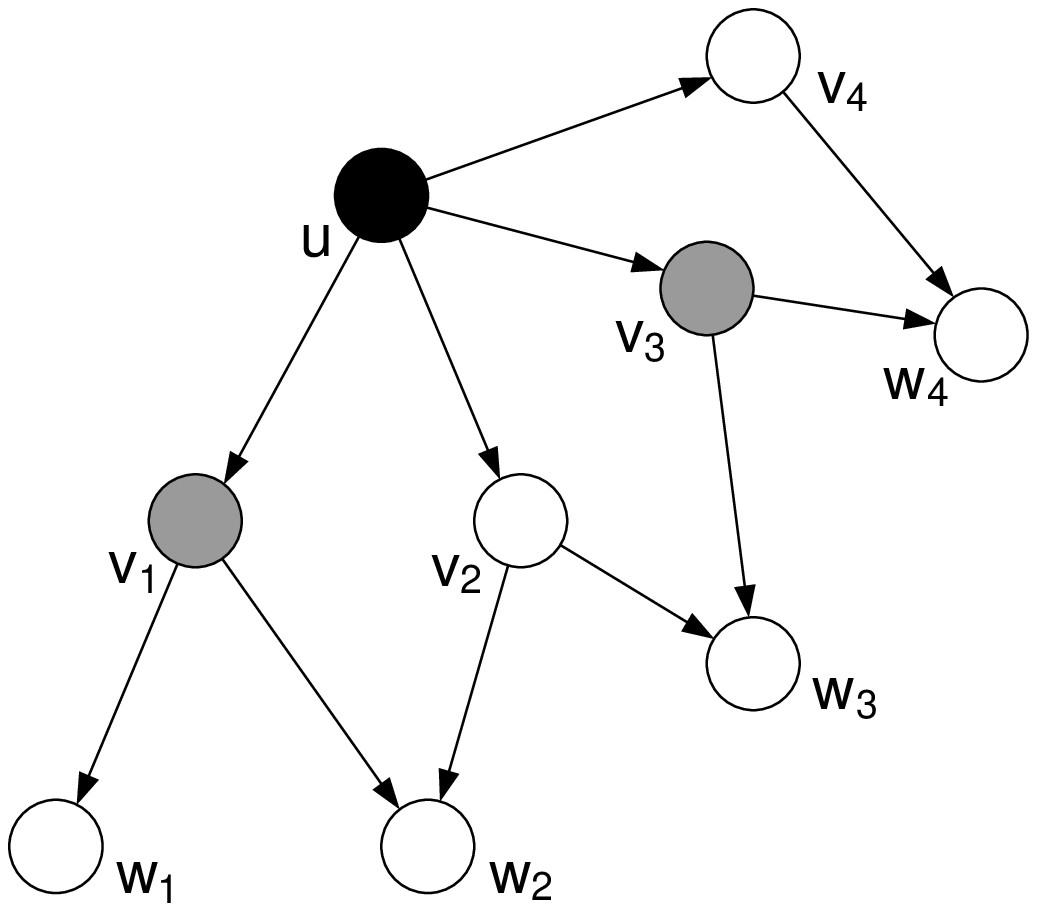}
\caption{Applying \mpr{} at node $u$: $\mbox{MPR}(u)=\{v_1,v_3\}$.}
\label{fig:mprExample}
\end{figure}

An example of this heuristic is given in Fig.~\ref{fig:mprExample}, starting with $\mbox{MPR}(u)=\emptyset{}$. The node $v_1$ is the only one able to reach $w_1$, so it is added to $\mbox{MPR}(u)$ and nodes $w_1$ and $w_2$ are removed from $\mbox{MPR}'(u)$. No other mandatory $1$-hop neighbor of $u$ exists, so other relays are selected according to the number of nodes in $\mbox{MPR}'(u)$ they cover. Nodes $v_2$ and $v_4$ cover only one node in $\mbox{MPR}'(u)$ , while node $v_3$ covers at the same time $w_3$ and $w_4$, so $v_3$ is chosen and added to $\mbox{MPR}(u)$. The set $\mbox{MPR}'(u)$ being empty, no other nodes are selected. We finally have $\mbox{MPR}(u)=\{v_1,v_3\}$.

\medskip

Being the broadcast protocol used in \olsr{}, \mpr{} has been the subject of miscellaneous studies since its publication. For example in \cite{BMF-ANALYSIS-AMPRSOLSR}, authors analyze how relays are selected and conclude that almost $75\%$ of them are selected in the first step of the greedy heuristic, so that improving the second step is not really useful. This conclusion seems correct, as long as the unit disk graph model is used.

 %%%%%%%%%%%%%%%%%%%%%%%%%%%%%%%%%%%%%%%%%%%%%%%%%%%%%
%   ____      _      _           __  __  ______  ___  %
%  / __ \____(_)__ _(_)__  ___ _/ / /  |/  / _ \/ _ \ %
% / /_/ / __/ / _ `/ / _ \/ _ `/ / / /|_/ / ___/ , _/ %
% \____/_/ /_/\_, /_/_//_/\_,_/_/ /_/  /_/_/  /_/|_|  %
%            /___/                                    %
 %%%%%%%%%%%%%%%%%%%%%%%%%%%%%%%%%%%%%%%%%%%%%%%%%%%%%
\section{Original Greedy Heuristic}
\label{sec:originalmpr}

\subsection{Graphs generation}

In this section, we provide results about the performance of \mpr{} over our considered realistic physical layer, the lognormal shadowing model. We chose not to use a general purpose simulator in order to focus on the area of our study: we thus implemented algorithms and models in our own simulator, so that we had to decide how to generate `realistic' graphs considering the realistic model.

We chose to consider the method cited in Sec.~\ref{sec:preliminaries}, which is based on \hello{} messages. Neighborhood information is stored in a table which is regularly cleaned in order to remove too old entries. An entry is too old when the corresponding host has not signaled itself since a given amount of time, that we denote by $x$. Beacon messages are regularly sent by each host to signal itself. Let us denote by $y$ the time between two \hello{} messages (we have $x>y$). A node $u$ sees a neighbor $v$ if it has received at least one \hello{} message during the last $y$ seconds. The probability $\mbox{p}_n(u,v)$ for this event to occur is equal to:

\begin{equation}
\mbox{p}_n(u,v) = 1 - \overline{\mbox{p}(u,v)}^{\frac{x}{y}}.
\end{equation}

For each directional edge, a random number is thus drawn to determine if it exists. This way, when a node $u$ is aware of the existence of a neighbor $v$, it can decide to send messages to the latter. Of course, $u$ cannot be ensured that its messages will reach $v$.
%Fig.~\ref{fig:edge-model} illustrates this model with $x=1$ and $y=3$.
We can easily conclude that long edges have a high probability to be unidirectional while short edges have a high probability to be bidirectional.

% \begin{figure}[t]
% \centering
% \includegraphics[angle=-90,width=\gnuplotFigWidth]{fig/edge-model.eps}
% \caption{Our realistic physical layer and its impact on edges.}
% \label{fig:edge-model}
% \end{figure}
% 
% \medskip

All the results were obtained with the following parameters. The network is static and always composed of $500$ nodes randomly distributed in a uniform manner over a square area whose size is computed in order to obtain a given average density. Edges are created using the method previously described, and for each measure, we took the average value obtained after $500$ iterations. We fixed the communication radius to be equal to $75$ in both physical models. An ideal \mac{} layer is considered to isolate the intrinsic properties of the selected relays: collisions of packets could skew both results and analyses.

\subsection{Experimental results}

\begin{figure}[t]
\centering
\subfigure[Receiving nodes.]{\label{subfig:perfMPRDiffusion} \includegraphics[angle=-90,width=\gnuplotFigWidth]{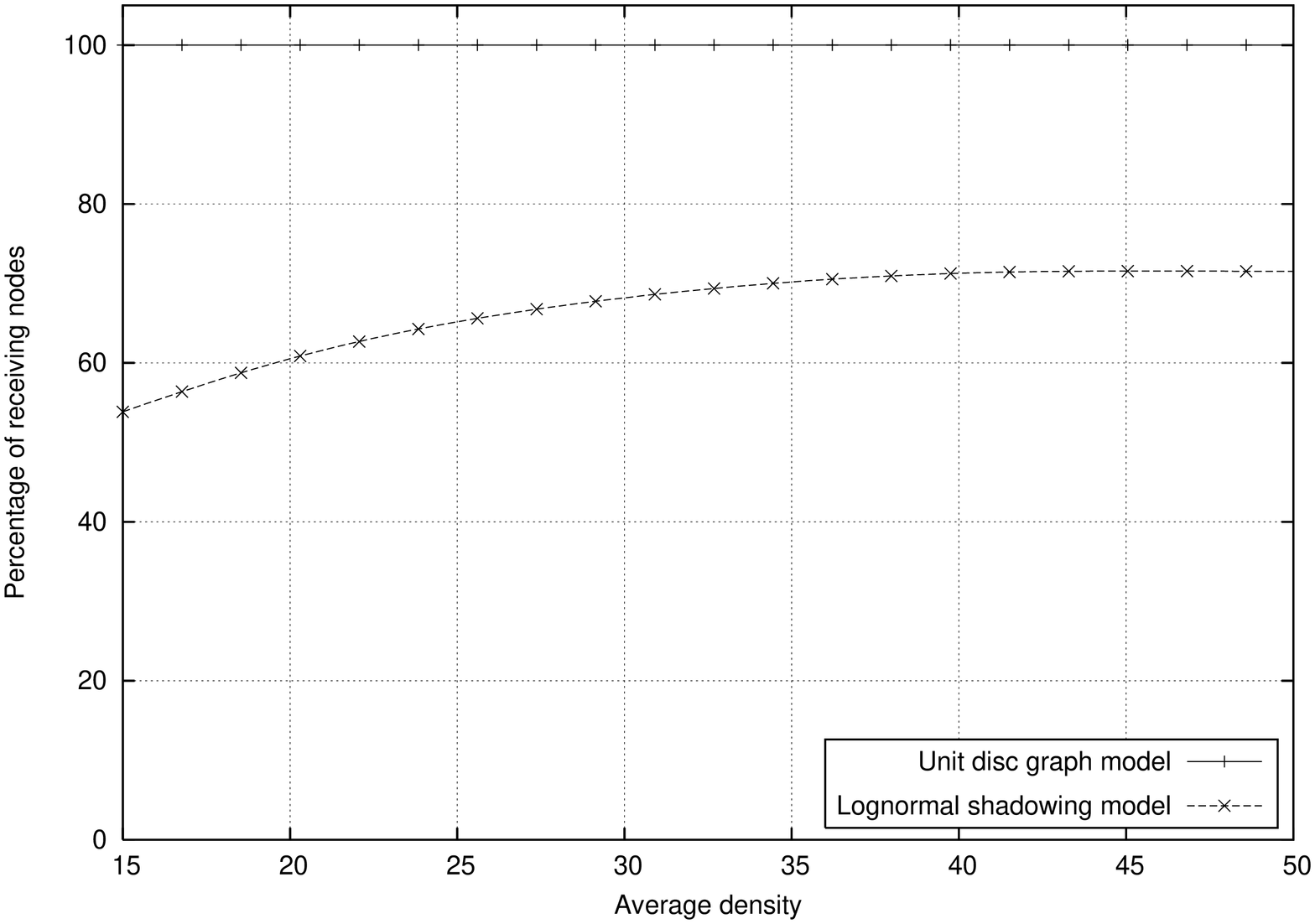}} \hfil
\subfigure[Transmitting nodes.]{\label{subfig:perfMPRRelays} \includegraphics[angle=-90,width=\gnuplotFigWidth]{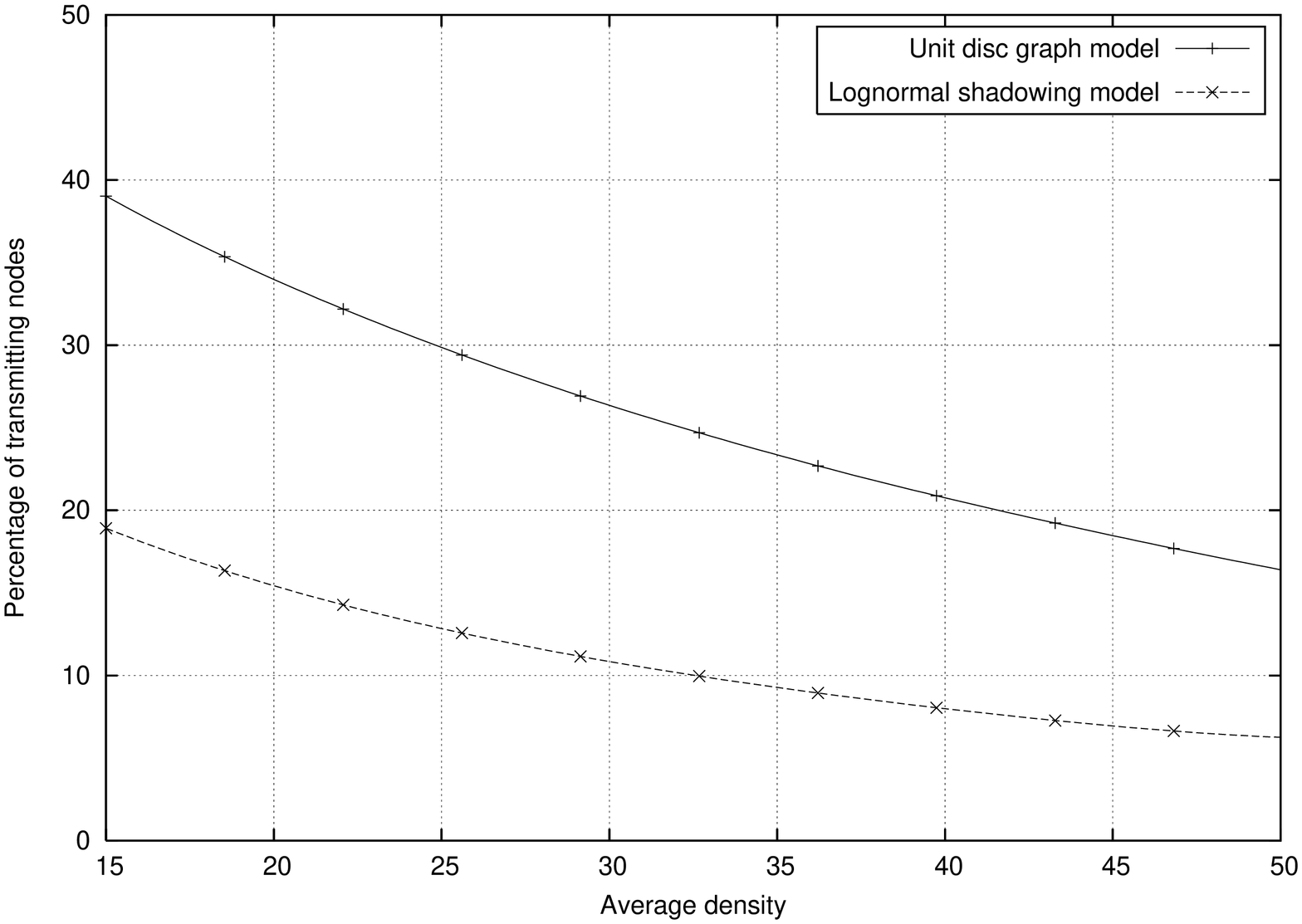}} \hfil
\caption{Performance of \mpr{} with the two considered physical models.}
\label{fig:perfMPR}
\end{figure}

We provide in Fig.~\ref{subfig:perfMPRDiffusion} the delivery ratio of \mpr{} using the two considered physical layers. When using the unit disk graph model, a total coverage of the network is achieved as \mpr{} is a deterministic algorithm. However, this is no more the case with the lognormal shadowing model due to the multiple errors of transmission: the delivery ratio is under $70\%$ for each considered density, and is as low as $55\%$ for a density $d=15$.

\begin{figure}
\centering
\includegraphics[angle=-90,width=\gnuplotFigWidth]{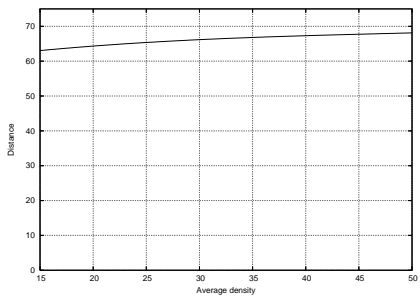}
\caption{Average distance between a node and its relays.}
\label{fig:distanceMPR}
\end{figure}

This poor performance can be explained by the fact that, as highlighted by Busson et al. in \cite{BMF-ANALYSIS-AMPRSOLSR}, the chosen relays are located at the limit of the communication range, where the probability of correct reception is low. This is confirmed in our experiments, as illustrated by Fig.~\ref{fig:distanceMPR}: the average distance between a node and its multipoint relays is almost equal to $68$, while the maximal communication range is $75$. Moreover, \cite{BMF-ANALYSIS-AMPRSOLSR} also states that $75$\% of the relays are chosen during the first step: this means that, when a relay does not correctly receive the message, there is a risk of $75$\% that this relay was the only one able to reach an isolated node, which will thus not receive the message, potentially leading to a partition of the network.

We also provide in Fig.~\ref{subfig:perfMPRRelays} the percentage of nodes which correctly received and then relayed the message. It is interesting to note that this percentage is different with the two models. Indeed, as only nodes which received the message are taken into account, one would have expected to observe the same values in both cases. This means that the needed number of relaying nodes does not linearly vary with the number of covered nodes: obviously, only a few relays are needed to cover a high number of different nodes, but a larger number is needed to cover the last few remaining ones.

 %%%%%%%%%%%%%%%%%%%%%%%%%%%%%%%%%%%%%%%%%%%%%%%%%%%%%%%%%%%
%    _  __             __ __             _     __  _        %
%   / |/ /__ _    __  / // /__ __ ______(_)__ / /_(_)______ %
%  /    / -_) |/|/ / / _  / -_) // / __/ (_-</ __/ / __(_-< %
% /_/|_/\__/|__,__/ /_//_/\__/\_,_/_/ /_/___/\__/_/\__/___/ %
%                                                           %
 %%%%%%%%%%%%%%%%%%%%%%%%%%%%%%%%%%%%%%%%%%%%%%%%%%%%%%%%%%%
\section{New Heuristics for \mpr{}}
\label{sec:newmpr}

As illustrated in the previous section, the original greedy heuristic used by Quayyum et al. in \cite{QVL-MULTIPOINT-MRFBMMWN} is not suitable for a realistic physical layer. An average delivery of $70\%$ is indeed not sufficient for most of applications, and an alternative solution must thus be used.

In this section, we propose miscellaneous replacement heuristics in order to improve the performance of \mpr{}. They aim at maximizing the average coverage, while minimizing the number of needed relays (and thus the energy consumption). In all our proposals, the first step of the original heuristic which allows isolated $2$-hop neighbors to be covered is kept (it is mandatory), only the second step is replaced.

We keep notations introduced in Sec.~\ref{sec:relatedwork}. Thus, considering a node $u$, the set $\mbox{MPR}(u)$ contains the multipoint relays chosen by $u$, while the set $\mbox{MPR}'(u)$ contains $2$-hop neighbors of $u$ not yet covered.

\subsection{First proposal: Straightforward approach}

As previously explained, the low delivery ratio of \mpr{} is caused by the too high distance between a node and its relays. The latter having little chance to correctly receive the broadcast packet, they also have little chance to be able to relay this packet and thus to cover the $2$-hop neighbors of the emitter.

A first and straightforward idea could be, when choosing a relay, to balance the coverage it offers and its probability to correctly receive the packet. Thus, at each step considering a node $u$, a score can be computed for each potential relay $v$. The node with the highest score is selected and placed in $\mbox{MPR}(u)$. We denote by $\mbox{c}_u(v)$ the \emph{additional} coverage offered by $v$ to $u$:

\begin{equation}
\mbox{c}_u(v) = |\mbox{MPR}'(u) \cap \mbox{N}(v)|.
\end{equation}

The score obtained by $v$ at a given iteration for a node $u$, denoted by $\mbox{s}_u(v)$, is thus defined by:

\begin{equation}
\mbox{s}_u(v) = \mbox{c}_u(v) \times \mbox{p}(u,v) .
\end{equation}

In simple terms, the additional coverage offered by $v$ is weighted by its probability to correctly receive the broadcast packet. In Fig.~\ref{fig:exampleMPRLNS}, the score $\mbox{s}_u(v_1)$ of $v_1$ is equal to $3 \times \mbox{p}(u,v_1)$.

\subsection{Second proposal: Clever approach}

The previous heuristic, while being more suitable for a realistic environment than the original one, still has an obvious flaw: it still takes into account additional coverage in a too simple way. One can thus easily imagine a situation where a very distant $1$-hop neighbor would offer an additional coverage such that the latter would compensate a low probability of correct reception. In this case, this neighbor would be selected as relay while its probability to correctly receive the packet, and thus to be able to relay it, would be very low. One can also imagine a situation where the distance between the relay and the $2$-hop neighbors it covers would be very high, such that the re-emission of this relay would have little chance to reach these $2$-hop neighbors.

\begin{figure}[t]
\centering
\includegraphics[width=85pt]{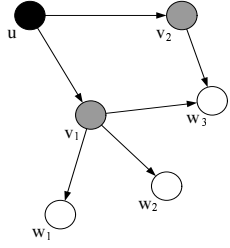}
\caption{A case where the node $u$ has to select its multipoint relays between its neighbors $v_1$ and $v_2$ ($\mbox{MPR}(u) = \emptyset{}, \mbox{MPR}'(u) = \{w_1, w_2, w_3\}$.}
\label{fig:exampleMPRLNS}
\end{figure}

\begin{figure*}
\centering
\subfigure[Receiving nodes.]{\label{subfig:mprnew-diffusion} \includegraphics[angle=-90,width=\gnuplotFigWidth]{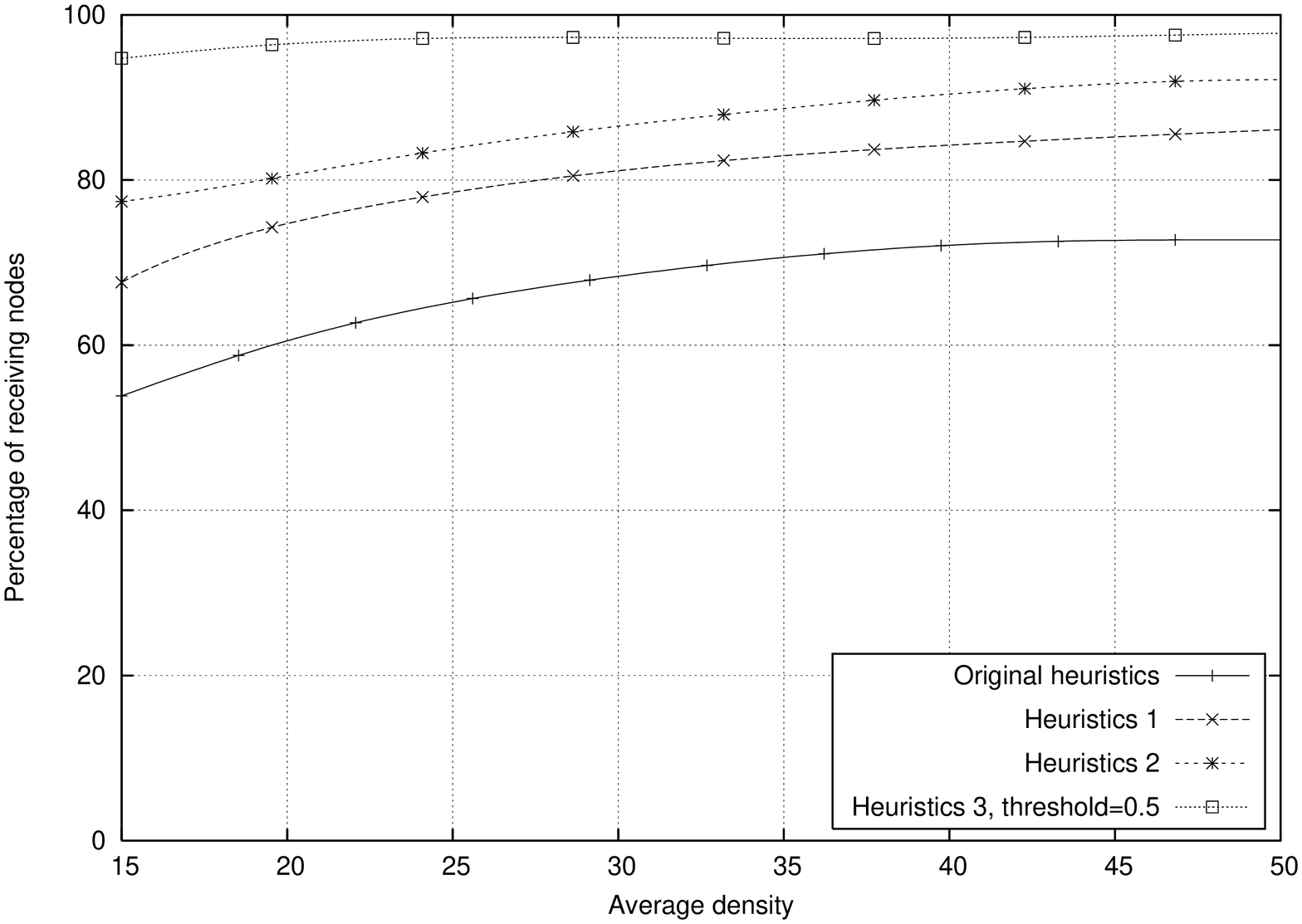}}   \hfil
\subfigure[Transmitting nodes.]{\label{subfig:mprnew-emitters} \includegraphics[angle=-90,width=\gnuplotFigWidth]{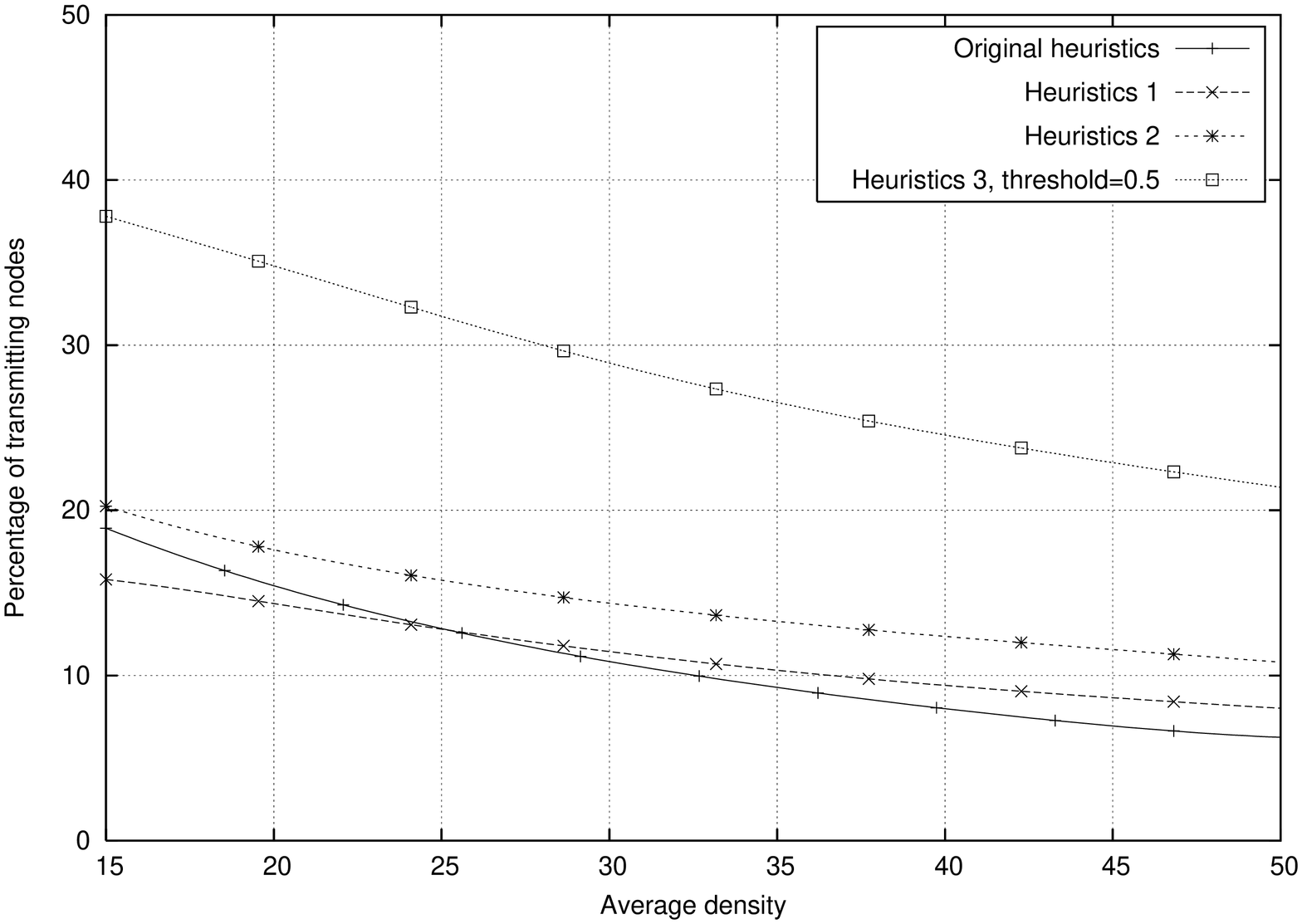}} \hfil
\caption{Performance of the different heuristics using the lognormal shadowing model.}
\label{fig:mprPerformancesNew}
\end{figure*}

We propose to extend the concept used in the first proposal, by taking into account the probabilities of correct reception between the potential relay and the $2$-hop neighbors it covers. We thus replace the additional coverage offered by a relay by the average probability of correct reception by $2$-hop neighbors. We thus obtain:

\begin{equation}
\mbox{s}_u(v) = \mbox{p}(u,v) \times \sum_{i=1}^{\;i=|\mbox{c}_u(v)|}(\;\mbox{p}(v,w_i) \, / \, |\mbox{c}_u(v)|\;).
\end{equation}

This way, multipoint relays offering a low coverage in terms of probabilities have little chance to be selected. In Fig.~\ref{fig:exampleMPRLNS}, the score $\mbox{s}_u(v_1)$ of $v_1$ is now equal to $\mbox{p}(u,v1) \times ( (\mbox{p}(v_1,w_1) + \mbox{p}(v_1,w_2) + \mbox{p}(v_1,w_3)) / 3)$.

\subsection{Third proposal: Robustness approach}

In the previous proposals, as soon as a $2$-hop neighbor has a non-null probability to be covered, it is removed from $\mbox{MPR}'(u)$. This removal is done even with a very low probability, which in this case may be meaningless. It can be more interesting to consider a $2$-hop neighbor as covered when its probability to correctly receive the broadcast packet is over a given threshold, in order to increase the delivery ratio.

We thus propose to keep the score computation used in the previous heuristic, while modifying how $2$-hop neighbors are removed from $\mbox{MPR}'(u)$. For such a $2$-hop neighbor $w$ of $u$, its removal from $\mbox{MPR}'(u)$ is done only if its coverage level $\mbox{t}_u(w)$ is over a given threshold. The value of $\mbox{t}_u(w)$ is given by:

\begin{equation}
\mbox{t}_u(w) = 1 - \prod_{i=1}^{i=|\mbox{MPR}(u)|} \overline{\mbox{p}(v_i,w)},
\end{equation}

\noindent
$\overline{\mbox{p}(v_i,w)}$ being equal to $1 - \mbox{p}(v_i,w)$. In simpler terms, the coverage level of a $2$-hop neighbor is equal to its probability to correctly receive the packet from at least one of the chosen relays.

Still considering Fig.~\ref{fig:exampleMPRLNS}, if the nodes $v_1$ and $v_2$ are selected as relays, then the coverage level $\mbox{t}_u(w_3)$ of $w_3$ is equal to $1 - (\overline{\mbox{p}(v_1,w_3)} \times \overline{\mbox{p}(v_2,w_3)})$. Several relays can thus now be selected to cover the same set of $2$-hop neighbors, in order to increase the delivery ratio. 

\begin{figure*}
\centering
\subfigure[Receiving nodes.]{\includegraphics[angle=-90,width=\gnuplotFigWidth]{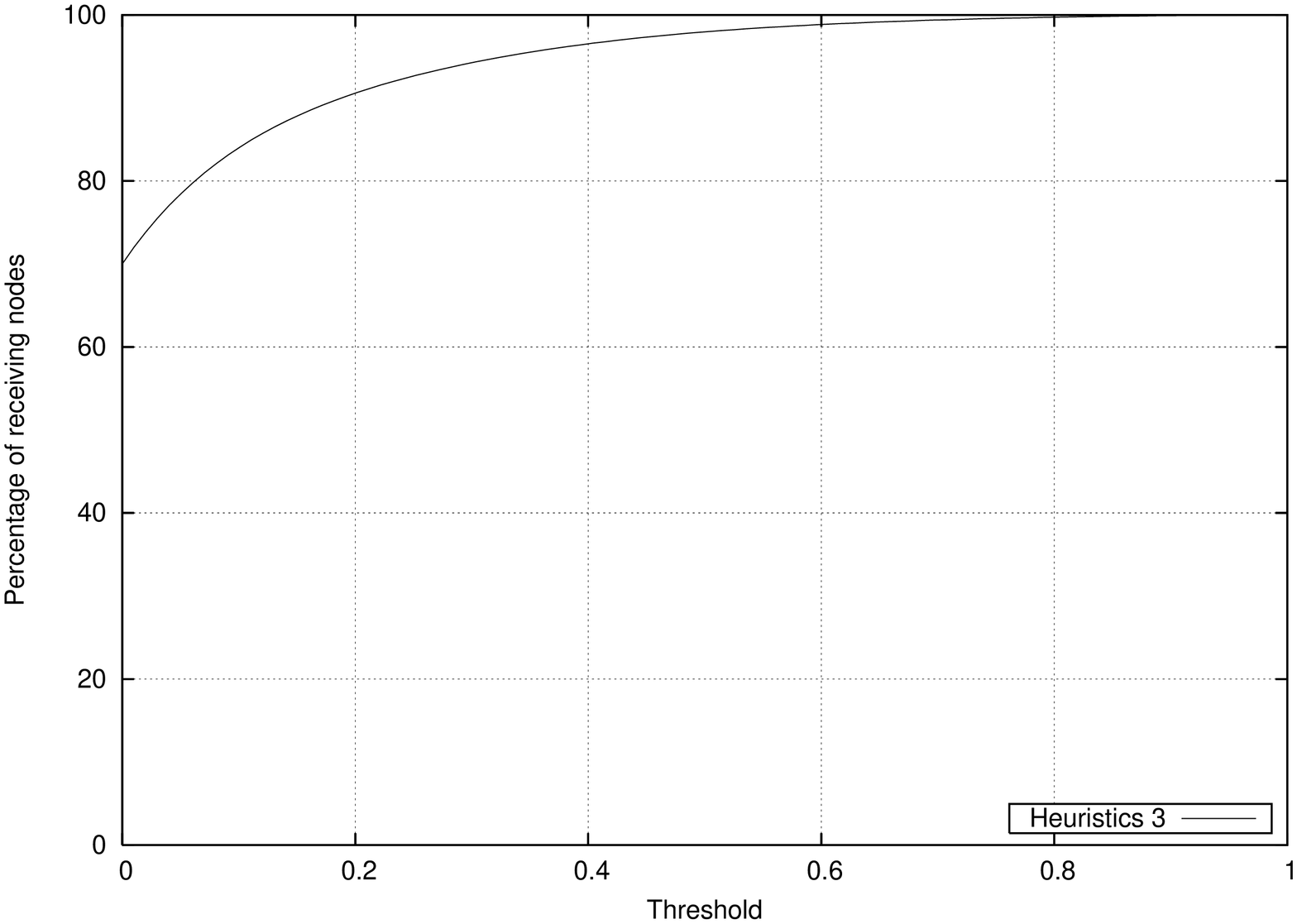}} \hfil
\subfigure[Transmitting nodes.]{\includegraphics[angle=-90,width=\gnuplotFigWidth]{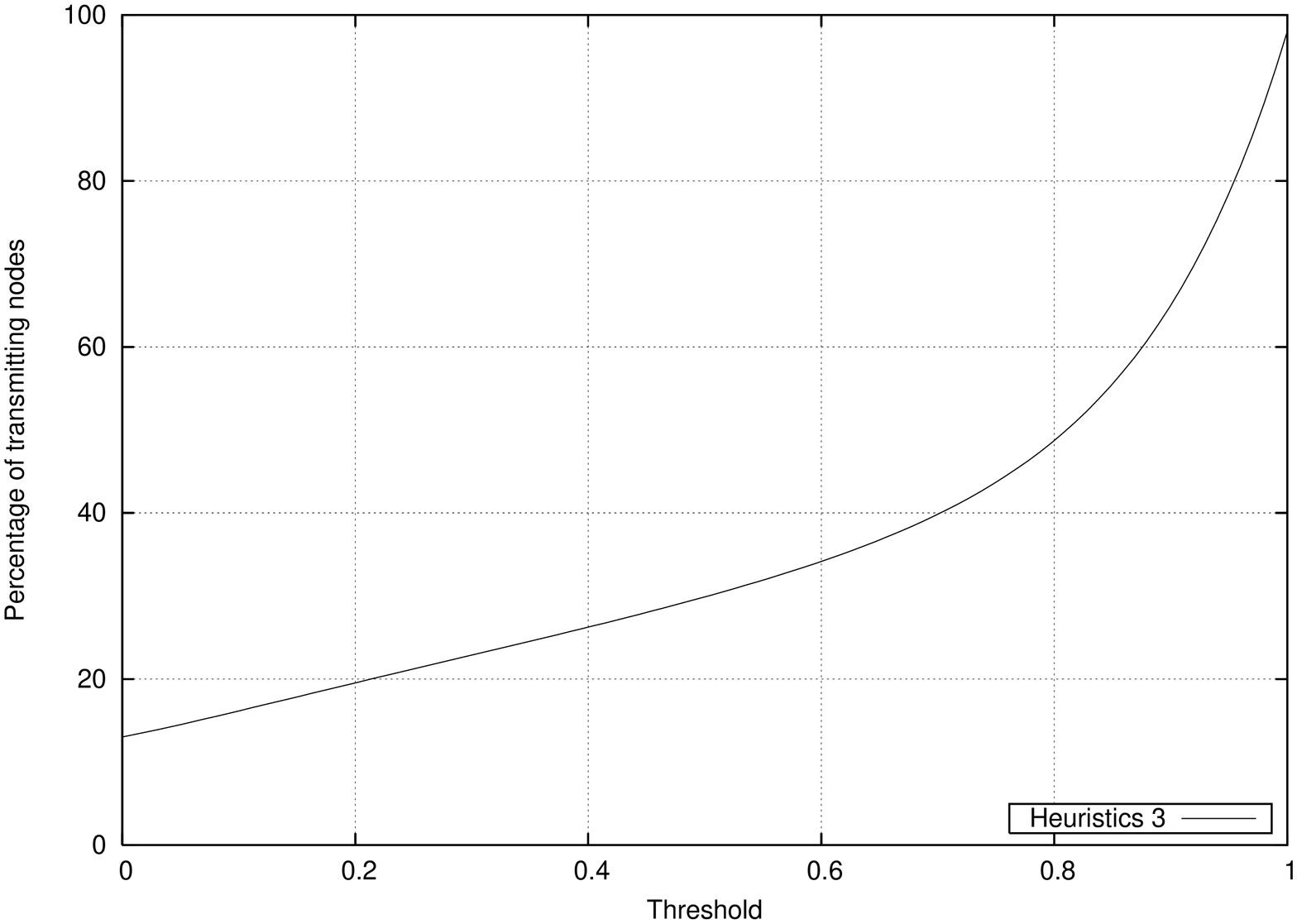}} \hfil
\caption{Performance of the third heuristic for varying thresholds and a density $d=30$.}
\label{fig:mprPerformancesHeuristics3}
\end{figure*}

\subsection{Performance}

We provide in Fig.~\ref{fig:mprPerformancesNew} the performance of the new heuristics presented in this section, considering the lognormal shadowing model. We use the same parameters as in Sec.~\ref{sec:originalmpr}.

Not surprisingly, we observe in Fig.~\ref{subfig:mprnew-diffusion} that the new heuristics lead to a far better delivery ratio than the original algorithm. This improvement is of course due to the use of the probabilities of correct reception given by the physical model. As expected, the second heuristic offers a higher percentage of covered nodes simply because it prevents too far neighbors to be selected as relays. Considering the density $d=30$, the original heuristic only covers $67$\% of nodes, against $81\%$, $85\%$ and $98\%$ for our three proposals. The delivery ratio has thus been greatly improved, as aimed by our heuristics.

As illustrated by Fig.~\ref{subfig:mprnew-emitters}, the third heuristic, used with a threshold equal to $0.5$, requires the participation of $28\%$ of the receiving nodes for the density $d=30$ to provide a delivery ratio of $98\%$. This may seem a high value compared to other curves, but considering the results given in Fig.~\ref{fig:perfMPR} with the unit disk graph model, one can observe that values are almost the same for the original heuristic. This means that the number of chosen multipoint relays for a given node is approximately the same, but their choice is of better quality.

We finally provide in Fig.~\ref{fig:mprPerformancesHeuristics3} the performance of the third heuristic for different values of the threshold parameter, considering a density $d=30$. As expected, the delivery ratio is proportional to the value of the threshold while the number of relays is inversely proportional to it. Choosing a threshold equal to $1$ is almost useless as a total coverage can nearly be achieved with a value between $0.4$ and $0.5$ with far less relaying nodes. Using a threshold of $0$ does not lead to a null delivery ratio, because the first step is still applied to cover isolated nodes.

 %%%%%%%%%%%%%%%%%%%%%%%%%%%%%%%%%%%%%%%%%%%%
%   _____              __         _          %
%  / ___/__  ___  ____/ /_ _____ (_)__  ___  %
% / /__/ _ \/ _ \/ __/ / // (_-</ / _ \/ _ \ %
% \___/\___/_//_/\__/_/\_,_/___/_/\___/_//_/ %
%                                            %
 %%%%%%%%%%%%%%%%%%%%%%%%%%%%%%%%%%%%%%%%%%%%
\section{Conclusion}
\label{sec:conclusion}

From the variety of results presented, we can observe that a realistic physical layer leads to miscellaneous problems while broadcasting. The \mpr{} protocol is a good example: while being very efficient with the unit disk graph, its delivery ratio is not sufficient for most applications with a realistic model. While this study focused on \mpr{}, we believe that other main broadcasting methods, such as dominating sets, will exhibit the same flaws. However, some small modifications, which takes into account probabilities of correct reception, may correct these flaws. Thus, the new heuristics we presented for \mpr{} keep the principle of the protocol, only the selection process of multipoint relays is modified. The latter, while being approximately as many as in the original heuristic, are generally better chosen and provide a higher delivery ratio.

\medskip

More generally, a huge amount of work is left to be done about this subject. As previously stated, other well-known algorithms will probably need to be modified in order to provide correct performance. Other mechanisms, such as the neighbor elimination scheme \cite{SS-BROADCASTING-BAIWN}, may be of prime importance in the quest for the optimal tradeoff between robustness and efficiency. Other aspects of communications, such as neighborhood discovery protocols must also be studied and probably adapted to realistic environments.
%Indeed, almost all broadcast protocols assume a correct knowledge of the neighborhood, and incorrect knowledge leads to very bad behaviors.

% Bibliography 
% \bibliographystyle{splncs}
% \bibliography{ingelrest}

% that's all folks
\end{document}